\documentclass[journal]{IEEEtran}
\IEEEoverridecommandlockouts
\usepackage{algorithmic, algorithm,lipsum}
\usepackage{psfrag}
\usepackage{amsthm,color}
\usepackage{cite}

\ifCLASSINFOpdf
   \usepackage[pdftex]{graphicx}
\else
   \usepackage[dvips]{graphicx}
\fi
\usepackage[cmex10]{amsmath}
\usepackage{amssymb}
\usepackage{array}

\usepackage{mdwmath}
\usepackage{mdwtab}

\usepackage{amsmath}

\newcommand{\beq}{\begin{equation}}
\newcommand{\eeq}{\end{equation}}

\begin{document}
%
\title{Massive MIMO with 1-bit ADC}

%

\author{ Chiara Risi, Daniel Persson, and Erik G. Larsson
\thanks{C. Risi is with the Department of Electrical and Information Engineering (DIEI), University of Cassino and Lazio Meridionale, 03043 Cassino (FR), Italy (E-mail: chiara.risi@unicas.it).}
\thanks{D. Persson and E. G. Larsson are with the Department of Electrical Engineering (ISY), Link\"oping University, 581 83 Link\"oping, Sweden 
(E-mail: danielp@isy.liu.se, erik.g.larsson@liu.se).}
}

\markboth{SUBMITTED TO THE IEEE TRANSACTIONS ON COMMUNICATIONS}{Shell \MakeLowercase{\textit{et al.}}: Bare Demo of IEEEtran.cls for Journals}


\maketitle

\begin{abstract}
We investigate massive multiple-input-multiple-output (MIMO) uplink systems with 1-bit analog-to-digital converters (ADCs) on each receiver antenna.
Receivers that rely on 1-bit ADC do not need energy-consuming interfaces such as automatic gain control (AGC).
This decreases both ADC building and operational costs.
Our design is based on maximal ratio combining (MRC), zero-forcing (ZF), and least squares (LS) detection, taking into account the effects of
the 1-bit ADC on channel estimation.

Through numerical results, we show good performance of the system in terms of mutual information and symbol error rate (SER). 
Furthermore, we provide an analytical approach to calculate the mutual information and SER of the MRC receiver. The analytical approach
reduces complexity in the sense that a symbol and channel noise vectors Monte Carlo simulation is avoided.

\end{abstract}


\IEEEpeerreviewmaketitle

\section{Introduction}
MIMO systems have attracted significant research interest during the last decade, and are incorporated into emerging wireless broadband standards like Long-Term Evolution (LTE) \cite{LTE}. 
In order to perform the receive processing, the received analog baseband signal is converted into digital form using a couple of analog-to-digital converters per antenna, i.e., one sampler each for the
in-phase and quadrature components. 

There are several types of ADC. One ADC type is the flash. It consists of $2^b$ comparators, where $b$ is the ADC resolution in bits. 
The receive voltage is divided over a resistive ladder with comparators measuring over different parts of the ladder. 
The comparators whose thresholds are less than their fed voltage give a non-zero output signal. These measurements are then transformed to bits.
An important part of the flash ADC is the automatic gain control, which amplifies the received signal so as to match it to the range of the resistor ladder and comparators.  
Other ADC architectures are pipelined ADC and sigma-delta ADC \cite{sigmadelta}.

Irrespectively of ADC technology, more output bits requires more operational power.
There are several MIMO studies that take into account the effects of the ADC on the performance evaluation of the system. 
The paper \cite{Murray} examines the ADC effects with a ZF filter at the receiver, while \cite{Mezghani2} and \cite{Mezghani4}
explore adaptation of the linear minimum mean square error (MMSE) receiver and the non-linear MMSE-decision feedback receiver 
to take into account the ADC presence. Maximum likelihood  detection with ADC is investigated in \cite{Mezghani3},
while \cite{Shah-Dabeer} focuses on beamforming techniques to improve the performance of a system with low precision ADC. 

The special case of 1-bit ADC is particularly interesting, since
no AGC is needed. 
While 1-bit ADC is advantageous in terms of hardware complexity and energy consumption, 
 it generally has a severe impact on performance. 
Ultra-wideband MIMO systems with 1-bit ADC are studied in \cite{Mezghani}. Rayleigh fading MIMO channels with 1-bit ADC are studied in \cite{Mezghani5}.
In \cite{Nossek2}, an analysis of binary space-time block codes with optimum decoding is provided for systems with 1-bit receive signal quantization.
All the above MIMO ADC treatments express the performance in terms of bit-error-rate and/or mutual information.

Massive MIMO as systems, also known as very large MIMO and large-scale antenna systems, have base stations (BSs) equipped with several hundred antennas, 
which simultaneously serve many tens of terminals in the same time-frequency resource \cite{MaMi2}. 
To be more specific, we define massive MIMO  as a system with 
$M$ BS antennas and $K$ users, where the inequality $M\gg K\gg 1$ holds.
Massive MIMO systems are known to be able to average out channel noise and fading \cite{MaMi1,MaMi2,MaMi3,MaMi4,MaMi5,MaMi7-Hien}.
No investigations have been conducted to decide if massive MIMO systems could be used  to average out ADC noise as well.
All the contributions on receiver design for massive MIMO systems today assume that the receiver has access to received data with 
infinite precision.

\subsection{Our contribution}
This paper considers the uplink of a massive MIMO system employing 1-bit ADCs. 
The main difference between our work and the works cited above is that we consider the massive MIMO case, i.e., $M\gg K\gg 1$.
Our contributions are the following.
\begin{itemize}

\item A discussion of maximum a posteriori probability (MAP) channel estimation with 1-bit ADC is provided. We quantify the computational complexity to be exponential in $K$. 
Since this computational complexity is high in a setting with many users, we propose a sub-optimal LS-channel estimation approach.

\item We suggest MRC and ZF filters based on the LS channel estimate described under the previous point.
We further derive an LS detection filter, which is calculated directly from the uplink training sequences without relying on an intermediate 
channel estimate. 

\item We derive an analytical expression for the probability distribution of the MRC filter soft symbol estimates. 
Using this probability distribution, closed form expressions are developed for both the mutual information between the transmitted symbols and hard symbol estimates and the mutual information between the transmitted symbols and the soft symbol estimates with MRC.
The closed-form expression reduces computational complexity in the sense that a Monte Carlo simulation is avoided when computing the mutual information.

\item The proposed systems are evaluated by numerical experiments. Both mutual information and SER are investigated. Numerical evaluations of the mutual information and SER are also compared to the their closed form expressions.
We conclude that massive MIMO provides excellent SER and mutual information performance for  wide ranges of system parameters.
  
\end{itemize}
The rest of the paper is organized as follows. In Section \ref{sec:1},  we describe the system model, detection filters, 
 channel estimation methods, and  analyze the performance of the system. Our proposed techniques are explored by experiments in Section \ref{sec:simulations}. The paper is concluded in Section \ref{sec:conclusions}.  

\medskip

\noindent\emph{Notation: } 
We use boldface lowercase and uppercase letters to denote vectors and matrices respectively.

\section{Massive MIMO uplink with 1-bit ADC and no AGC} \label{sec:1}
This section describes the channel model, suggested solutions, and performance analysis. The system model is detailed in Section \ref{sec:system_model}, and Section \ref{sec:ch_estimation} describes the employed detection filters and channel estimation methods.
Numerical and analytical procedures for estimating the mutual information are discussed  in Section \ref{sec:analysis}.

\subsection{System model} \label{sec:system_model}
We consider a single-cell uplink, where there are $K$ single-antenna users and one BS equipped with an array of $M$ antennas.
The discrete-time complex baseband received signal at the base station is
\begin{equation}
\mathbf{r}=\sqrt{P_t}\mathbf{H}\mathbf{x}+\mathbf{n}, \label{y_complex}
\end{equation}
where  $\mathbf{H}\in\mathbb{C}^{{M}\times {K}}$ is the channel matrix between the BS and the $K$ users, i.e., $h_{ij}$ is the channel coefficient between the $j$-th user and the $i$-th antenna of the BS.
The entries of $\mathbf{H}$ are independent $\mathcal{CN}(0,1)$ random variables. 
The vector $\mathbf{x}\in\mathbb{C}^{K}$ contains the transmitted symbols from all the users. In particular, the $j$-th entry of $\mathbf{x}$,  $x_{j}$, is the symbol transmitted by user $j$. The symbols are modeled as independent identically distributed random variables with zero mean and variance $\operatorname{E}[\mid x_{j} \mid^2]=1$ and, since the spectral efficiency in a massive MIMO system is typically low \cite{MaMiSE}, we assume that they belong to a QPSK constellation. 
Finally, $P_t$ is the transmit power per user, and $\mathbf{n} \in  \mathbb{C}^{K}$ is the noise vector. The entries of $\mathbf{n}$ are independent identically distributed 
zero-mean circularly symmetric complex Gaussian random variables, which we denote by $\mathcal{CN}(0,\sigma_N^2)$.

%
Let $Q(\cdot):\mathbb{R}\rightarrow \{1,-1\}$ represent the 1-bit ADC quantizer function. It is defined as  $\operatorname{Q}(u)=\operatorname{sign}(u)$, 
where $u\in \mathbb{R}$ and $\operatorname{sign}(u)$ is the sign function that returns $1$ if $u \geq 0$ , and $-1$ if $u<0$. For a complex number $v=v_R+jv_I$, $Q(\cdot)$ is applied separately for the real and imaginary parts as
$\operatorname{Q}(v)=\operatorname{sign}(v_R)+j\operatorname{sign}(v_I)$, and for a vector it is applied element-wise.
The quantized received signal is expressed as
\begin{equation}
\mathbf{y}=\operatorname{Q}(\mathbf{r}).\label{y_q}
\end{equation}
A soft estimate of the transmitted symbols is obtained by processing the quantized received vector through the receive filter $\mathbf{A}$ as follows
\begin{equation}
\mathbf{\widetilde{x}}=\mathbf{A}^H\mathbf{y}.\label{x_soft}
\end{equation}
Finally, it is possible to perform a QPSK demodulation that gives as output a hard estimate $\widehat{\mathbf{x}}$ of the transmitted vector $\mathbf{x}$.

\subsection{Receive filters} \label{sec:ch_estimation} 
In order to derive the soft symbol estimates, the receive filter $\mathbf{A}$ needs to be chosen. 
The channel information needed to compute this filter is acquired through sending uplink pilots. 
In particular, each coherence time of the channel is divided in two parts. During the first part of it, the users transmit pilot symbols from which the base station calculates the receive filter. 
During the remainder of the coherence time, all the users transmit their data to the BS, which in turn uses the receive filter calculated from the uplink pilots to detect the symbols.
We assume 1-bit analog-to-digital conversion even during the reception of the pilots.

To compute the receive filter $\mathbf{A}$ directly as function of the training sequences, we use a LS approach. The filter obtained via the LS approach is defined as 
\begin{align}
&\mathbf{A}^H=
\underset{\mathbf{\widetilde{A}}}{\operatorname{argmin}} \left( \frac{1}{N}\sum_{n=1}^N \bigg|\bigg|
\mathbf{\widetilde{A}}^H \mathbf{y}^{(n)} - \mathbf{x}^{(n)} \bigg|\bigg|^2 \right)\notag \\
&=\left(\sum_{n=1}^N \mathbf{x}^{(n)} \left({\mathbf{y}^{(n)}}\right)^H \right)\left(\sum_{n=1}^N \mathbf{y}^ {(n)} \left({\mathbf{y}^{(n)}}\right)^H \right)^{-1}. \label{LS_filter}
\end{align}
In (\ref{LS_filter}), $N$ is the number of time instances used for the pilot transmission, while $\mathbf{x}^{(n)}$ and $\mathbf{y}^{(n)}$ are the transmitted vector and the quantized received vector, respectively, at the $n$-th time instance dedicated to the training transmission.
Note that the matrix $\sum_{n=1}^N \mathbf{y}^{(n)} \left({\mathbf{y}^{(n)}}\right)^H$ can be invertible only if $N\geq M$.
As a consequence, the LS filter can only be used in a scenario where the channel conditions vary slowly. 
The LS computational complexity grows linearly with the number of users $K$ and cubicly with the number of antennas at the BS $M$.
We have derived the LS receive filter directly from the pilot symbols. Another possibility is to derive an estimate of the channel matrix $\mathbf{H}$ first, and then calculate the filter based on the channel state information (CSI). 

For MAP-optimal channel estimation, we calculate
\begin{align}
\mathbf{\widehat{H}}&=
\underset{\mathbf{H}}{\operatorname{argmax}} \;  p(\mathbf{H}|\mathbf{Y})=
\underset{\mathbf{H}}{\operatorname{argmax}} \; p(\mathbf{Y}|\mathbf{H})p(\mathbf{H}), \label{MAP}
\end{align}
where $\mathbf{Y}=[\mathbf{y}^{(1)},...,\mathbf{y}^{(N)}]$, $\mathbf{X}=[\mathbf{x}^{(1)},...,\mathbf{x}^{(N)}]$, $p(\mathbf{H}|\mathbf{Y})$ is the probability density function (pdf) of $\mathbf{H}$ given $\mathbf{Y}$, $p(\mathbf{Y}|\mathbf{H})$ is the probability mass function (pmf) of $\mathbf{Y}$ given $\mathbf{H}$, and $p(\mathbf{H})$ is the pdf of $\mathbf{H}$.
By independence of the variables involved at different antennas, the optimization problem in (\ref{MAP}) is equivalent to 
\begin{equation}
\mathbf{\widehat{h}}_i=\underset{\mathbf{h}_i}{\operatorname{argmax}}  \; p\left(\mathbf{Y}_i |\mathbf{h}_i \right)p(\mathbf{h}_i),  \: i=1,...,M,
\label{MAP_single}
\end{equation}
where $\mathbf{\widehat{h}}_i^T$, $\mathbf{h}_i^T$, and $\mathbf{Y}_i^T$ are the $i$-th rows of $\mathbf{\widehat{H}}$, $\mathbf{H}$, and $\mathbf{Y}$ respectively, $p\left(\mathbf{Y}_i|\mathbf{h}_i \right)$ is the pmf of $\mathbf{Y}_i$ given $\mathbf{h}_i$, and $p(\mathbf{h}_i)$ is the pdf of $\mathbf{h}_i$.
In order to solve the $i$-th optimization problem in (\ref{MAP_single}), we have to resort to a grid search method. In particular, the space $\mathbb{C}$ is discretized so that $h_{ij} $ belongs to a finite set $\mathcal{A} \; \forall i,j $. 
The BS has to search over all possible vectors $\mathbf{h_i}$, and there are $\lvert \mathcal{A} \rvert^K$ such  vectors. Therefore, MAP estimation has a computational complexity that is exponential in the number of users. 
A main benefit of massive MIMO is to be able to multiplex many users, i.e., $K\gg 1$, and we can conclude that MAP channel estimation is less suitable for this application.

To reduce the computational complexity, we use the LS estimator. The LS estimate of $\mathbf{H}$ is given by
\begin{align}
&\mathbf{\widehat{H}}=
\underset{\mathbf{\widetilde{H}}}{\operatorname{argmin}} \sum_{n=1}^N \bigg|\bigg| 
 \mathbf{y}^{(n)} - \sqrt{P_t} \mathbf{\widetilde{H}} \mathbf{x}^{(n)} \bigg|\bigg|^2 \notag \\
&=\left(\sum_{n=1}^N \sqrt{P_t} \mathbf{y}^{(n)} {\mathbf{x}^{(n)}}^H \right)\left(\sum_{n=1}^N P_t \mathbf{x}^{(n)} {\mathbf{x}^{(n)}}^H \right)^{-1}. \label{channel_estimation}
\end{align}
Note that, in this case, we need at least $K$ training symbol transmissions for the matrix $\sum_{n=1}^N  \mathbf{x}^{(n)} {\mathbf{x}^{(n)}}^H$ to be invertible.
The computational complexity, instead, grows linearly with the number of antennas at the BS $M$ and cubicly with the number of users $K$. 
The computational complexity reduction with respect to (\ref{LS_filter}) is thus significant. Lastly, we refer to $\mathbf{\widehat{H}}=\mathbf{H}$ as the case of full CSI.

The next step is to derive the receive filter $\mathbf{A}$ based on the LS $\mathbf{H}$-estimate. We consider two conventional linear detectors, namely the MRC and the ZF.
The MRC is defined as
\begin{equation}
\mathbf{a}^i=\frac{{\mathbf{\widehat{h}}^i}}{\parallel{\mathbf{\widehat{h}}^i}\parallel^2},
\end{equation}
where $\mathbf{a}^i$ and $\mathbf{\widehat{h}}^i$ are the $i$-th columns of $\mathbf{A}$ and $\mathbf{\widehat{H}}$, respectively.
The ZF is defined as
\begin{equation}
\mathbf{A}^H=\mathbf{\widehat{H}}^\dagger,
\end{equation}
where $\mathbf{\widehat{H}}^\dagger=\left(\mathbf{\widehat{H}}^H \mathbf{\widehat{H}}\right)^{-1} \mathbf{\widehat{H}}^H$ is the pseudoinverse of the matrix $\mathbf{\widehat{H}}$.\\

\subsection{Performance metrics}\label{sec:analysis}
In this section, we consider two performance metrics: the mutual information per user and the SER per user. We show how they can be calculated numerically and analytically. 

\subsubsection{Mutual information} \label{MI}
The average mutual information of the discrete channel between the transmitted QPSK symbol $x_k$ and the QPSK symbol estimate $\widehat{x}_k$ is defined as
\begin{align}
&\operatorname{I}(x_k;\widehat{x}_k)=  \operatorname{E}_\mathbf{H} \left[ \sum_{x_k,\widehat{x}_k} p(\widehat{x}_k \rvert x_k ,\mathbf{H}) p(x_k) \log_2 \frac{p(\widehat{x}_k\rvert x_k, \mathbf{H})}{p(\widehat{x}_k \rvert \mathbf{H})}\right],\notag \\ 
& \label{MI_numerical}
\end{align}
where ${p(\widehat{x}_k\rvert x_k, \mathbf{H})}$ is the pmf of $\widehat{x}_k$ given $x_k$ and $\mathbf{H}$, $p(x_k)$ is the pmf of a QPSK constellation with uniformly distributed symbols, and 
$p(\widehat{x}_k \rvert \mathbf{H})= \sum_{{x_k}}  p(\widehat{x}_k\rvert {x_k},\mathbf{H})p({x_k})$. \\
\textit{In a case where the mutual information is evaluated numerically, ${p(\widehat{x}_k\rvert x_k, \mathbf{H})}$ is estimated through Monte Carlo simulations 
with many realizations of the transmit vector $\mathbf{x}$ and the channel noise vector $\mathbf{n}$.
In what follows, we give an alternative to the Monte Carlo evaluation of ${p(\widehat{x}_k\rvert x_k, \mathbf{H})}$. }

To calculate the transition probabilities ${p(\widehat{x}_k\rvert x_k,\mathbf{H})}$, we need to find first $f(\widetilde{x}_k\rvert x_k,\mathbf{H})$, which is the pdf of $\widetilde{x}_k$ given $x_k$ and $\mathbf{H}$.
Considering a matched filtering, note that $\widetilde{x}_k$ can be rewritten as
\begin{align} 
\widetilde{x}_k &=\sum_{i=1}^M \frac{h_{ik}^*}{\lVert \mathbf{h}_k \rVert^2 } y_i\notag \\
&= \sum_{i=1}^M \left(\frac{h_{ik}^*}{\lVert \mathbf{h}_k \rVert^2 } y_i -\mu_i + \mu_i\right)\notag \\
&=\sum_{i=1}^M \left(  \frac{h_{ik}^*}{\lVert \mathbf{h}_k \rVert^2 } y_i -\mu_i \right)+ \sum_{i=1}^M \mu_i, \label{x_soft_MRC}
\end{align}
where $\mu_i=\operatorname{E}_{y_i}\left[\frac{h_{ik}^*}{\lVert \mathbf{h}_k \lVert^2 } y_i \big| x_k, \mathbf{H} \right]=  \frac{h_{ik}^*}{\lVert \mathbf{h}_k \rVert^2 } \sum_{c=1}^4 p_{ic} s_c$, $p_{ic}= \operatorname{Prob}\left(y_i=s_c\rvert x_k, \mathbf{H}\right)$ and $s_1=1+j$, $s_2=1-j$, $s_3=-1-j$, $s_4=-1+j$.

In order to calculate ${p(\widehat{x}_k\rvert x_k,\mathbf{H})}$, we use the Cramer's central limit theorem \cite{Cramer,CentralTheorem}, which states that the sum $\sum_{i=1}^L u_i$ of a large number of independent
 random variables is approximately complex Gaussian if the following conditions are satisfied:
\begin{enumerate}
\item every component $u_i$ has a zero mean value,
\item every component $u_i$ has a finite variance $a_i^2=\operatorname{E}[\lvert u_i\rvert^2]$,
\item $\frac{a_i}{s_L}\xrightarrow{L\rightarrow \infty} 0$ and $s_L \xrightarrow{L\rightarrow \infty} \infty$, where $s_L=\sum_{i=1}^L a_i^2$.
\end{enumerate}
 
To apply the Cramer's central limit theorem to (\ref{x_soft_MRC}), we define $z_i =\left[\left(  \frac{h_{ik}^*}{\lVert \mathbf{h}_k \rVert^2 } y_i -\mu_i \right)| x_k, \mathbf{H} \right]$, and we make Assumption (A) that $z_i$ and $z_j$ are independent for all $i,j$ such that $i\neq j$.
\textit{Further in this section, we will show that, with Assumption (A), the approximated pdf of $\widetilde{x}_k$ derived below fits well with the real probability distribution,
 and in Section \ref{sec:simulations}, we will show that the performance evaluation based on the approximated pdf closely match the Monte Carlo symbol and noise vectors simulation results}.\\
Applying the Cramer's central limit theorem with (\ref{x_soft_MRC}) and Assumption (A) above, we have that 
\begin{equation}
\widetilde{x}_k \rvert x_k ,\mathbf{H} \xrightarrow{M\rightarrow \infty} \mathcal{CN}\left(\sum_{i=1}^M \mu_i, \sum_{i=1}^M \sigma_i^2 \right), \label{pdf_x_soft}
\end{equation}
where $\sigma_i^2= \frac{|h_{ik}|^2}{\lVert \mathbf{h}_k \rVert^4 } \left( 2-\lvert \sum_{c=1}^4 p_{ic} s_c \rvert^2\right)$.
In order to calculate $p_{ic}$, we consider
\begin{equation}
y_i=\operatorname{sign}\left(\overbrace{ \sqrt{P_t} h_{ik} x_k+ \underbrace{\sqrt{P_t}\sum_{j=1, j\neq k}^K h_{ij} x_j}_I + n_i }^{r_i}\right).
\end{equation}
Rewriting $I$ as
\begin{align}
I&=\sqrt{P_t} \sum_{j=1, j\neq k}^K h_{ij} x_j\notag\\
&=\sqrt{P_t} \sum_{j=1, j\neq k}^K ( h_{ij} x_j -\mu^{\rm{I}}_{j}) +\sqrt{P_t} \sum_{j=1, j\neq k}^K \mu^{\rm{I}}_{j}, \label{interference}
\end{align}
with $\mu^{\rm{I}}_{j}=\operatorname{E}_{{x}_j}\left[h_{ij}x_j | h_{ij}\right]=0$, we can again apply  the  Cramer central limit theorem on $\sum_{j=1, j\neq k}^K ( h_{ij} x_j -\mu^{\rm{I}}_{j})$ and find that   
\begin{align}
&r_i \rvert x_k, \mathbf{H} \xrightarrow{K\rightarrow \infty} \mathcal{CN}\left(\sqrt{P_t}h_{ik} x_k ,\underbrace{ P_t \sum_{j=1, j\neq k}^K \arrowvert h_{ij} \arrowvert^2}_{\left({\sigma^{\rm{I}}}\right)^2} + \sigma_N^2 \right).\notag \\
& \label{pdf_yi}
\end{align} 
The superscript $(\cdot)^I$  of the parameters $\mu^{\rm{I}}_{j}$ and $\left({\sigma^{\rm{I}}}\right)^2 $ is used to indicate that they are related to the interference term in (\ref{interference}).\\
Noticing that the real and imaginary parts of $r_i$ given $x_k$ and $\mathbf{H}$ are independent, and that they have the same variance, the probabilities $p_{ic}$ can be calculated using complementary error functions. In particular, for $x_k=\frac{1}{\sqrt{2}}+j\frac{1}{\sqrt{2}}$ they are derived as follows
\begin{align}
p_{i1}&=\operatorname{Prob} \left(y_i=1+j \Big\rvert x_k=\frac{1}{\sqrt{2}}+j\frac{1}{\sqrt{2}},\mathbf{H}\right)\notag \\
&= \frac{1}{2} \operatorname{erfc} \left(-\frac{\sqrt{P_t }\operatorname{Re} (h_{ik}x_k)}{\sqrt{\sigma_N^2 + \left({\sigma^{\rm{I}}}\right)^2} }\right)  \notag \\
& \;\;\cdot \frac{1}{2} \operatorname{erfc} \left(-\frac{\sqrt{P_t }\operatorname{Im} (h_{ik}x_k)}{\sqrt{\sigma_N^2 + \left({\sigma^{\rm{I}}}\right)^2} }\right), \label{erf_1}
\end{align}
\begin{align}
p_{i2}&=\operatorname{Prob} \left(y_i=1-j \Big\rvert x_k=\frac{1}{\sqrt{2}}+j\frac{1}{\sqrt{2}},\mathbf{H}\right)\notag \\
&= \frac{1}{2} \operatorname{erfc} \left(-\frac{\sqrt{P_t }\operatorname{Re} (h_{ik}x_k)}{\sqrt{\sigma_N^2 + \left({\sigma^{\rm{I}}}\right)^2}}  \right) \notag \\
&\;\cdot \left(1-\frac{1}{2} \operatorname{erfc} \left(-\frac{\sqrt{P_t }\operatorname{Im} (h_{ik}x_k)}{\sqrt{\sigma_N^2 + \left({\sigma^{\rm{I}}}\right)^2}} \right)  \right),\label{erf_2}
\end{align}
\begin{align}
p_{i3}&=\operatorname{Prob} \left(y_i=-1-j \Big\rvert x_k=\frac{1}{\sqrt{2}}+j\frac{1}{\sqrt{2}},\mathbf{H}\right)\notag \\
&= \left(1-\frac{1}{2} \operatorname{erfc} \left(-\frac{\sqrt{P_t }\operatorname{Re} (h_{ik}x_k)}{\sqrt{\sigma_N^2 + \left({\sigma^{\rm{I}}}\right)^2} }\right)  \right) \notag \\
&   \;\; \cdot \left(1-\frac{1}{2} \operatorname{erfc} \left(-\frac{\sqrt{P_t }\operatorname{Im} (h_{ik}x_k)}{\sqrt{\sigma_N^2 + \left({\sigma^{\rm{I}}}\right)^2}} \right)  \right),\label{erf_3}
\end{align}
\begin{align}
p_{i4}&=\operatorname{Prob} \left(y_i=1-j \Big\rvert x_k=\frac{1}{\sqrt{2}}+j\frac{1}{\sqrt{2}},\mathbf{H}\right)\notag \\
&= \left(1-\frac{1}{2} \operatorname{erfc} \left(-\frac{\sqrt{P_t }\operatorname{Re} (h_{ik}x_k)}{\sqrt{\sigma_N^2 + \left({\sigma^{\rm{I}}}\right)^2}} \right)  \right)\notag \\
&\quad \cdot \frac{1}{2} \operatorname{erfc} \left(-\frac{\sqrt{P_t }\operatorname{Im} (h_{ik}x_k)}{\sqrt{\sigma_N^2 + \left({\sigma^{\rm{I}}}\right)^2}} \right),  \label{erf_4}
 \end{align} 
 and similarly for other possible values of $x_k$.
Knowing how to calculate $p_{ic}$ as in (\ref{erf_1}) to (\ref{erf_4}), (\ref{pdf_x_soft}) now gives us the distribution of $\widetilde{x}_k $ given $x_k$ and $\mathbf{H}$.

\begin{figure*}
\centering \includegraphics[width=1\textwidth]{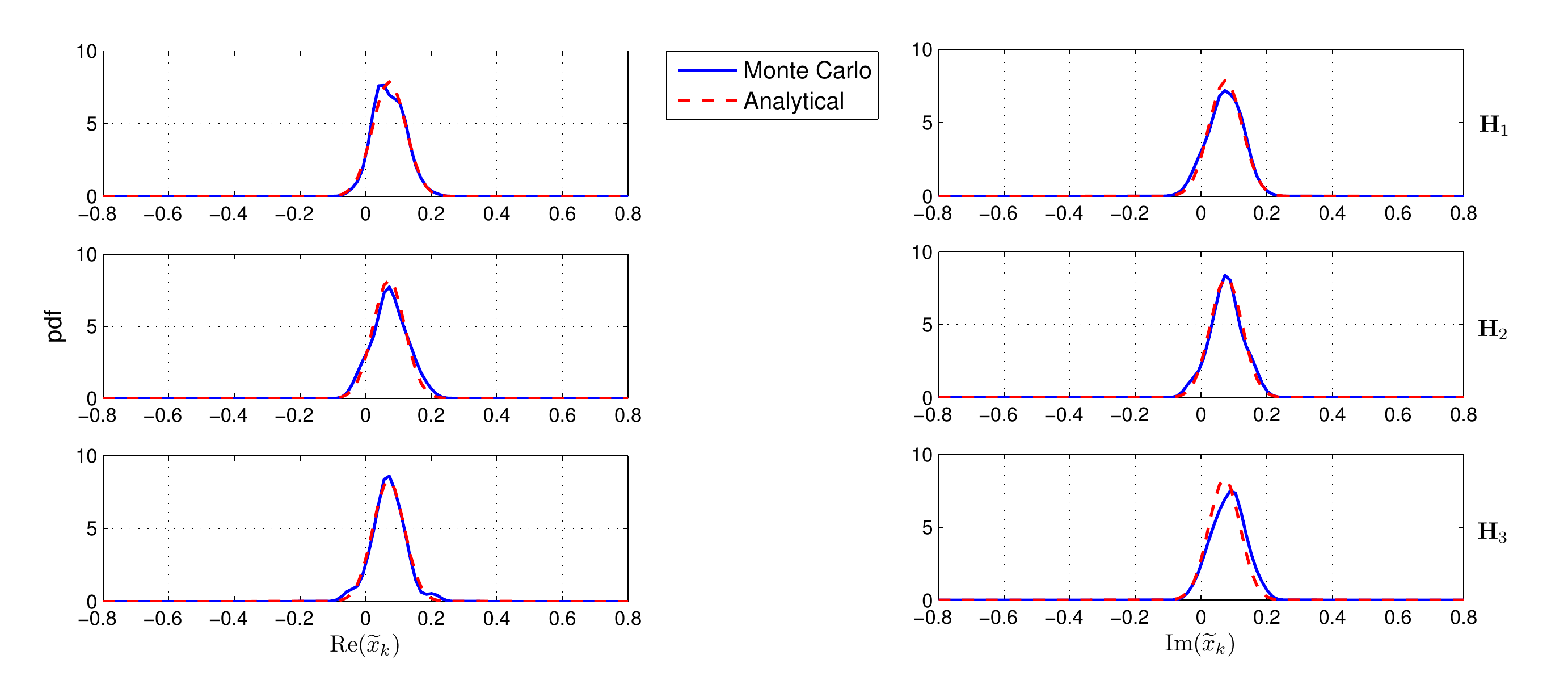} 
\caption{Probability distribution of the real and imaginary parts of $\widetilde{x}_k $ 
given $x_k=\frac{1}{\sqrt{2}}+j\frac{1}{\sqrt{2}}$ and $ \mathbf{H}$, considering $M=400$ and $K=20$. 
Three different channel matrices are considered, and the SNR is set to $-20$ dB.} \label{fig:gaussian_approximation}
\end{figure*}

In order to validate the model (\ref{pdf_x_soft}) derived using Assumption (A) for the soft symbol estimate, 
we provide the results in Figure \ref{fig:gaussian_approximation}. 
The plot is divided in a $3 \times 2 $ grid. The conditional probability distribution function of the real
 part of $\widetilde{x}_k$ given $x_k$ and $\mathbf{H}$ is depicted in the first column, while the distribution of the imaginary part of $\widetilde{x}_k$ given $x_k$ and $\mathbf{H}$ is depicted in the second column. The distributions are calculated both by Monte Carlo simulation and analytically, as in (\ref{pdf_x_soft}).
Further, each row corresponds to a different random channel realization. 
We consider a $400$-antenna BS that serves $K=20$ users.
The symbol $x_k$ transmitted by user $k$ is $\frac{1}{\sqrt{2}}+j\frac{1}{\sqrt{2}}$.
The signal-to-noise ratio (SNR) is set to $-20 $ dB, where the SNR is defined as $P_t/\sigma_N^2$. The results 
remain similar for all other SNR values, and for all tested matrix realizations. 
From Figure \ref{fig:gaussian_approximation}, we observe that the derived pdf in (\ref{pdf_x_soft}) closely matches the distribution obtained by Monte Carlo simulation.

Once that $f(\widetilde{x}_k\rvert x_k,\mathbf{H})$ is known, the transition probabilities ${p(\widehat{x}_k\rvert x_k,\mathbf{H})}$ can be calculated either numerically or using complementary error functions. Finally, the mutual information associated with the discrete channel between $x_k$ and $\widehat{x}_k$ is calculated as in (\ref{MI_numerical}).

In what follows, we also introduce the mutual information between the transmitted QPSK symbol $x_k$ and the soft symbol estimate $\widetilde{x}_k$. It is defined as
\begin{align}
&\operatorname{I}(x_k;\widetilde{x}_k)=\notag\\  &=\int \operatorname{E}_\mathbf{H}\left[ \sum_{x_k} f(\widetilde{x}_k\rvert x_k,\mathbf{H}) p(x_k) \log_2 \frac{f(\widetilde{x}_k\rvert x_k,\mathbf{H})}{f(\widetilde{x}_k\rvert \mathbf{H})} \right] d\widetilde{x}_k.
\end{align}
The integral over $\widetilde{x}_k$ is calculated by discretization of the continuous random variable $\widetilde{x}_k$. The resulting discrete random variable is $\widetilde{x}_k^\Delta$, and the mutual information between $\widetilde{x}_k^\Delta$ and $x_k$ is given by
\begin{align}
&\operatorname{I}(x_k;\widetilde{x}_k^\Delta)=\notag\\  &=\operatorname{E}_\mathbf{H}
\left[ \sum_{x_k,\widetilde{x}_k^\Delta} p(\widetilde{x}_k^\Delta \rvert x_k,\mathbf{H}) p(x_k) \log_2 \frac{p(\widetilde{x}_k^\Delta \rvert x_k,\mathbf{H})}{p(\widetilde{x}_k^\Delta\rvert \mathbf{H})} \right], \label{MI_soft}
\end{align}
where $p(\widetilde{x}_k^\Delta \rvert x_k,\mathbf{H})$ is the pmf of $\widetilde{x}_k^\Delta$ given  $x_k$ and $\mathbf{H}$,
 and $p(\widetilde{x}_k^\Delta\rvert \mathbf{H})=\sum_{{x_k}}  p(\widetilde{x}_k^\Delta\rvert {x_k},\mathbf{H})p({x_k})$. 
Note again that $p(\widetilde{x}_k^\Delta \rvert x_k,\mathbf{H})$ does not need to be estimated by Monte Carlo simulations based on many realizations of the transmit vector $\mathbf{x}$ and the channel noise vector $\mathbf{n}$. 
Instead we can rely on (\ref{pdf_x_soft}).

%

\subsubsection{SER}
The average SER of user $k$ is expressed as
\begin{align}
&\operatorname{SER}= p(\widehat{x}_k \neq x_k)\label{SER_numerical}\\
&=\operatorname{E}_\mathbf{H}\left[ \sum_{x_k} \sum_{\widehat{x}_k \neq x_k} p(\widehat{x}_k \rvert x_k,\mathbf{H})p(x_k) \right], \label{SER_analytical}
\end{align}
where $p(\widehat{x}_k \neq x_k)$ is the probability that the QPSK symbol estimate $\widehat{x}_k$ is different from the transmitted QPSK symbol $x_k$.\\
In a case where the SER is evaluated numerically, $p(\widehat{x}_k \neq x_k)$ in (\ref{SER_numerical}) is estimated through Monte Carlo simulations with many realizations of the transmit vector $\mathbf{x}$, the channel noise vector $\mathbf{n}$, and the channel matrix $\mathbf{H}$.\\
When the MRC filter is employed, we can alternatively use (\ref{SER_analytical}) to calculate the SER. The transition probabilities $ p(\widehat{x}_k \rvert x_k,\mathbf{H})$ in (\ref{SER_analytical}) are calculated as described in the previous section for the analytical evaluation of the mutual information. 

\section{Simulation results}\label{sec:simulations}

We will compare the proposed schemes in terms of Monte Carlo channel simulations.
If nothing else is mentioned, the results have been obtained numerically by Monte Carlo simulations, 
i.e., not by the analytical treatment employing (\ref{pdf_x_soft}).
Moreover, if nothing else is mentioned, 
the mutual information based on (\ref{MI_numerical}) associated with the 
discrete channel between the transmitted and received QPSK symbols $x_k$ and $\widehat{x}_k$ is depicted,
i.e.,  the mutual information per user between the transmitted QPSK symbol $x_k$ and the soft symbol estimate
$\widetilde{x}_k$ defined as in (\ref{MI_soft}) is not 
 depicted. 

Symbol and noise vectors Monte Carlo simulation mutual information is obtained by averaging the mutual information 
over $10^2$ channel matrix realizations, and for each channel realization, $10^2$ random symbol and noise vectors realizations are
 used. Symbol and noise vectors Monte Carlo simulation SER graphs, instead, are obtained by averaging over $10^5$  channel matrix, symbol and noise vectors realization triplets.
Analytic analysis curves based on (\ref{pdf_x_soft}) are obtained by averaging the mutual information over $10^2$ independent random channel matrix
realizations and the SER over $10^5$ channel random channel matrix realizations.  

\begin{figure}
\psfrag{SNR}[c]{SNR (dB)}
\centering \includegraphics[width=0.5\textwidth]{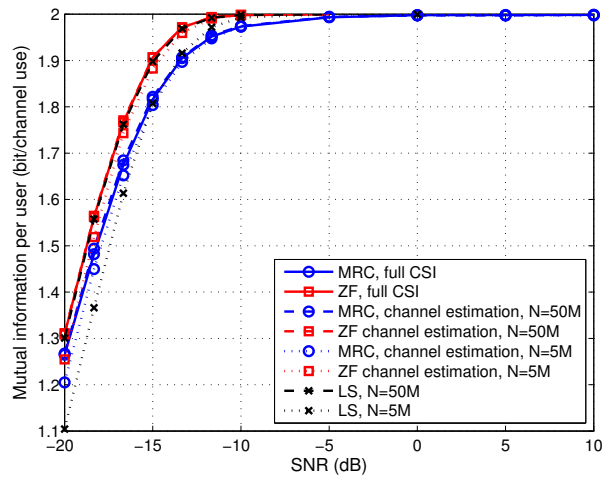} 
\caption{Mutual information per user based on (\ref{MI_numerical}) versus SNR for $M=400$ and $K=20$, and MRC, ZF, and LS filters. The MRC and ZF are considered for both the cases of perfect and imperfect CSI.} \label{fig:MIvsSNR_M=400_K=20_different_M}
\end{figure}

Figure \ref{fig:MIvsSNR_M=400_K=20_different_M} shows the mutual information per user  versus the SNR for MRC, 
ZF, and LS receivers, with $M=400$ antennas at the BS that serve $K=20$ users. 
The MRC and the ZF filters are investigated for both the cases of full CSI and imperfect channel knowledge. In order to calculate the 
channel estimate in (\ref{channel_estimation}) and the LS filter matrix in (\ref{LS_filter}), we choose pilot sequences of
length $N=5M$ and $N=50M$. The pilot sequences are randomly generated. It is seen that the LS filter requires around $N=50M$ 
time slots to achieve the same performance as the full CSI ZF filter. The MRC and ZF filters, instead, exhibit a faster convergence,
so that they require shorter training phase (the channel estimation error is almost suppressed using $N=5M$ time slots), as explained in Section \ref{sec:ch_estimation}. 

\begin{figure}
\centering \includegraphics[width=0.50\textwidth]{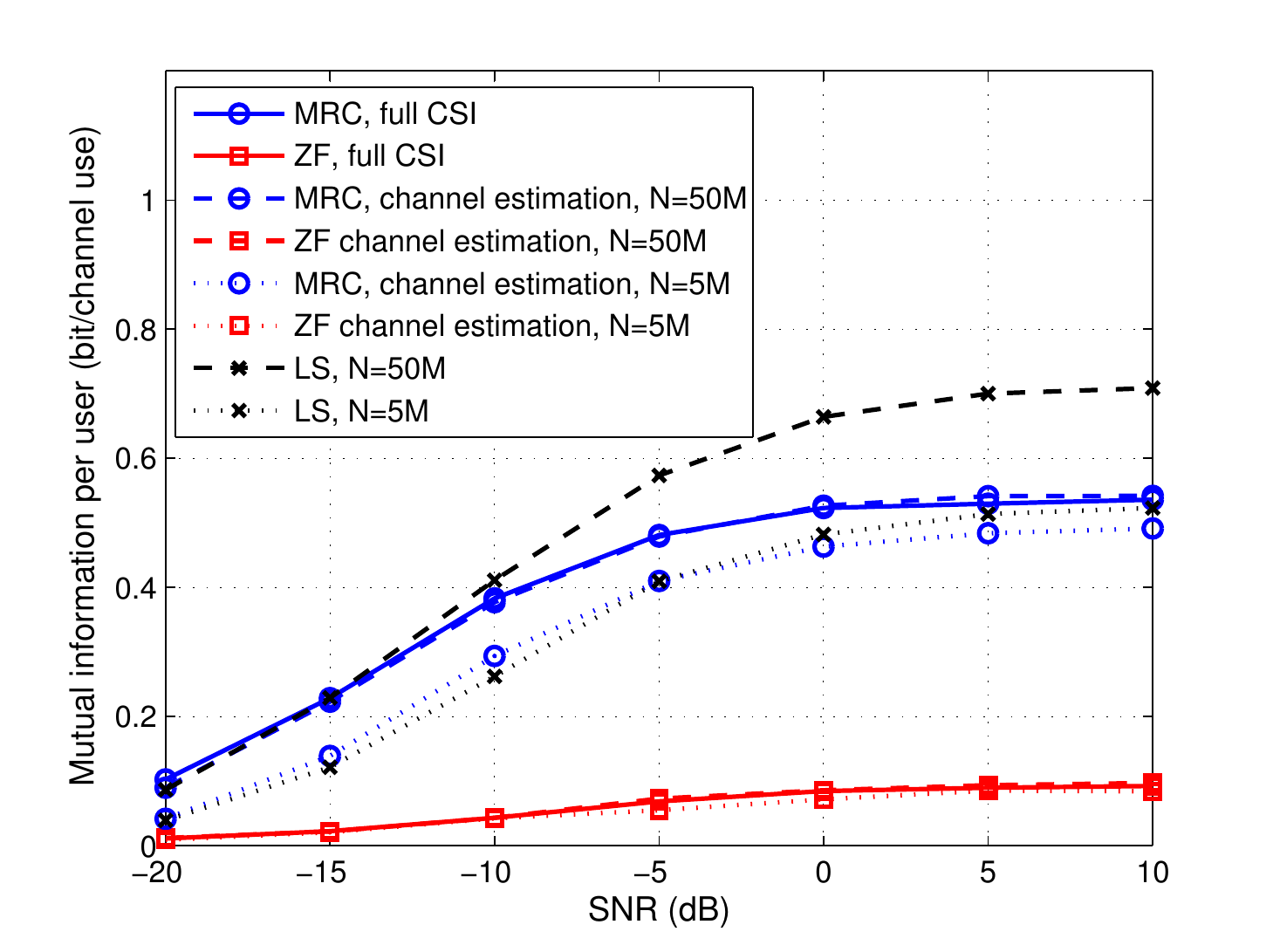} 
\caption{Mutual information per user based on (\ref{MI_numerical}) versus SNR for $M=20$ and $K=20$, and MRC, ZF, and LS filters. The MRC and ZF are considered for both the cases of perfect and imperfect CSI.} \label{fig:MIvsSNR_M=20_K=20_different_M}
\end{figure}

Figure \ref{fig:MIvsSNR_M=20_K=20_different_M} is organized in the same way, but considering a BS equipped with $M=20$ antennas and $K=20$ users.
Differently from the massive MIMO case, here the LS filter is performing better with respect to the ZF and MRC when $N=50M$. However, the 
performance of all methods is lower compared to that achieved by a massive MIMO system. Since the LS filter needs more training data to perform
well in the massive MIMO case, as is seen in Figure \ref{fig:MIvsSNR_M=400_K=20_different_M}, and  since it also needs higher computational complexity 
compared to the ZF and MRC channel estimation, as discussed in Section \ref{sec:ch_estimation}, we do not consider it in the following. 

\begin{figure}
\centering \includegraphics[width=0.5\textwidth]{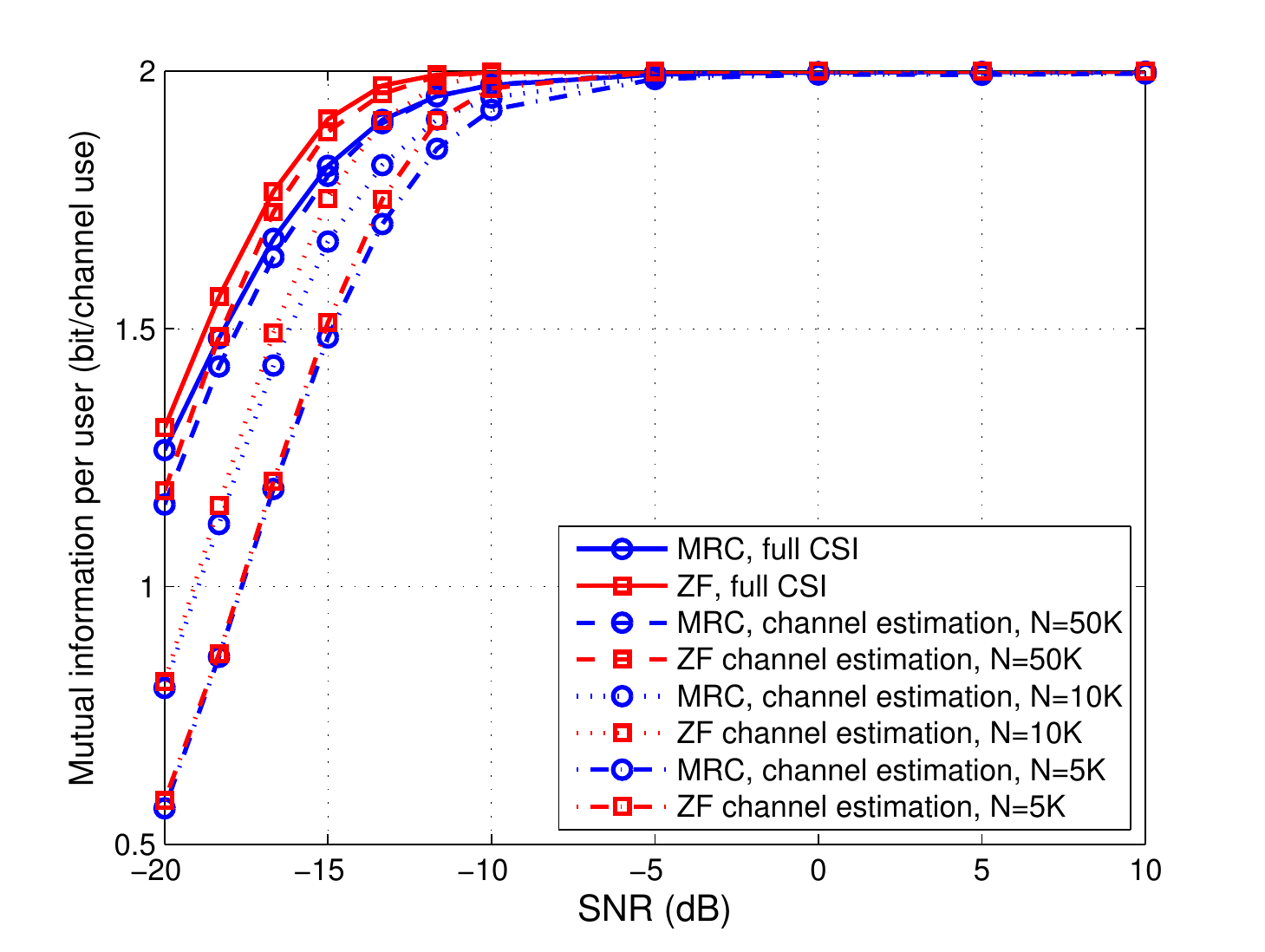} 
\caption{Mutual information per user based on (\ref{MI_numerical}) versus SNR for $M=400$ and $K=20$. The ZF and MRC filters are considered for both the cases of perfect and imperfect CSI. } \label{fig:MIvsSNR_different_N}
\end{figure}

\begin{figure}
\centering \includegraphics[width=0.5\textwidth]{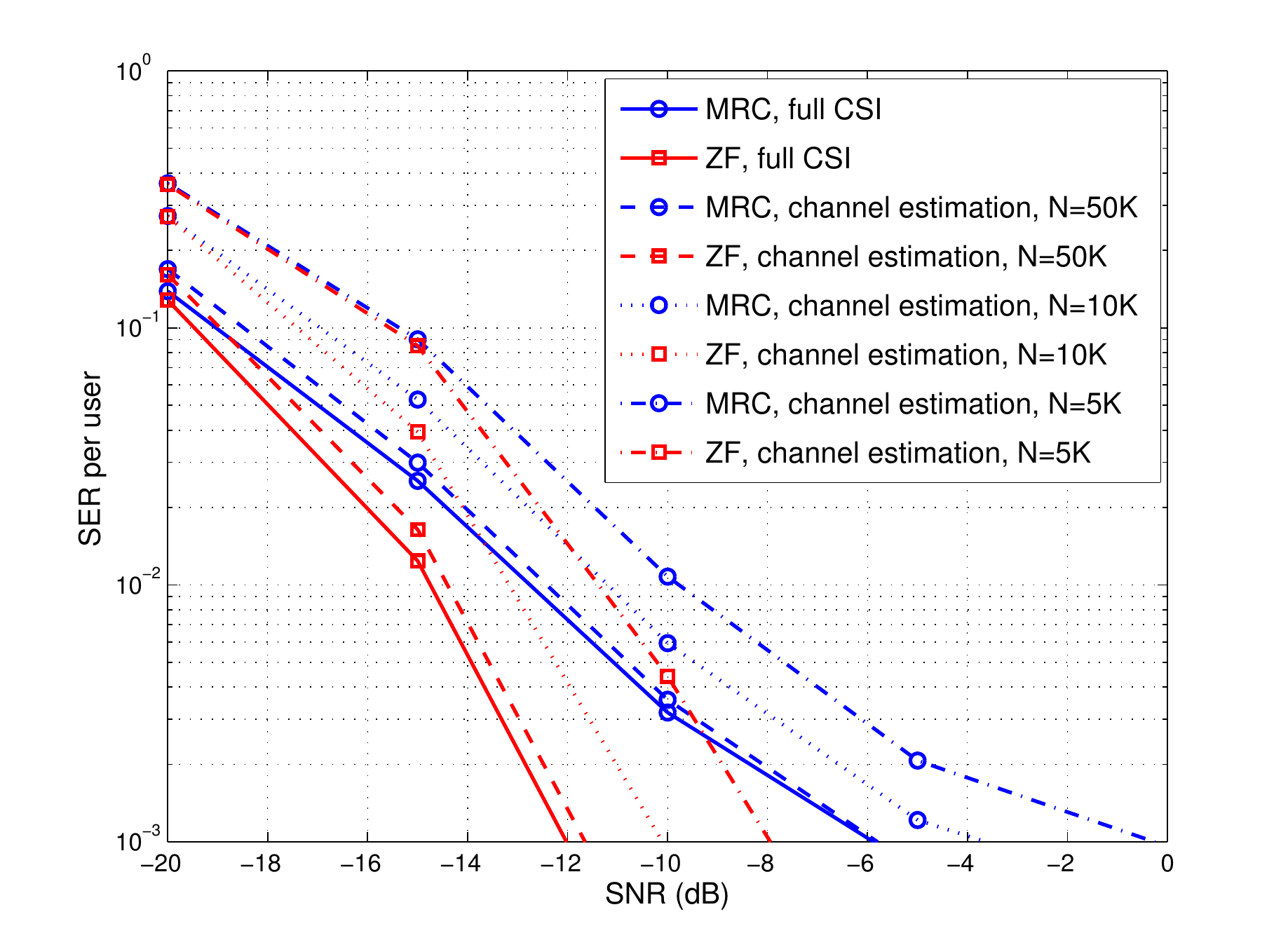} 
\caption{SER per user versus SNR for $M=400$ and $K=20$. The ZF and MRC filters are considered for both the cases of perfect and imperfect CSI. } \label{fig:SERvsSNR_different_N}
\end{figure}

Focusing the 
attention on the MRC and the ZF receivers, in Figure \ref{fig:MIvsSNR_different_N} and Figure \ref{fig:SERvsSNR_different_N} the mutual information
per user versus the SNR and the  SER per user versus the SNR  are depicted, respectively.
The aim of the graphs is to show how the length of the 
training sequences $N$ affects the performance. In particular, we consider three values of $N$: $5K$, $10K$, and $50K$. As we can see from the 
figures, to achieve the same performance as the full CSI case we have to use a long training sequence. This is possible in a scenario where 
the environment does not vary very fast. However, for values of SNR greater than $-5$ dB, the MRC and the ZF filters are equivalent in terms 
of mutual information, and the channel estimation error is suppressed even for short training sequences. In this same SNR regime, the maximal 
possible QPSK capacity of 2 bits is obtained.

\begin{figure}
\centering \includegraphics[width=0.5\textwidth]{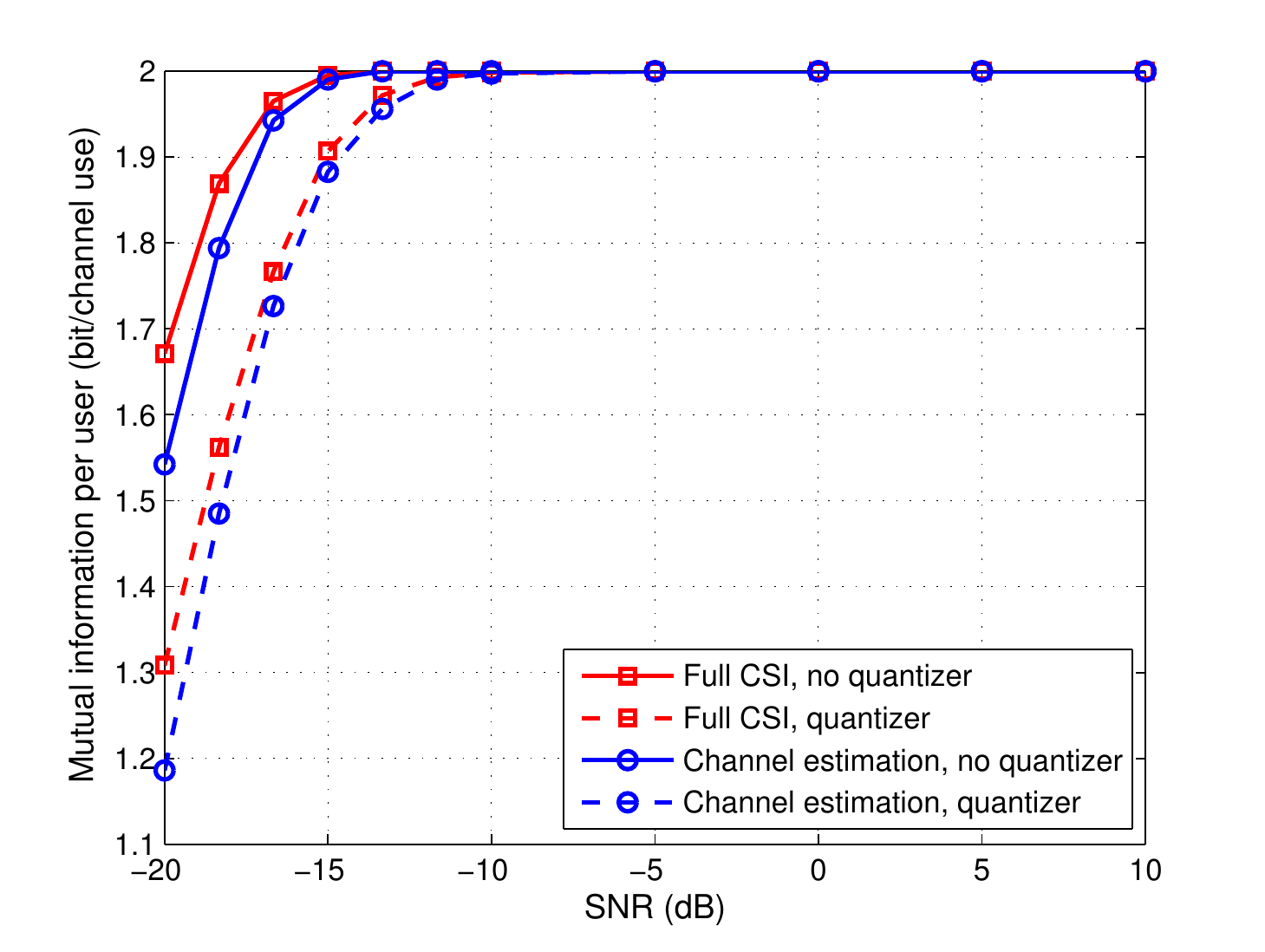} 
\caption{Mutual information per user based on (\ref{MI_numerical}) versus SNR considering $M=400$ and $K=20$, for the case in which the quantizer is taken into account and the case in which it is not. The ZF is considered for both the cases of perfect and imperfect CSI.} \label{fig:MIvsSNR_gap}
\end{figure}

\begin{figure}
\centering \includegraphics[width=0.5\textwidth]{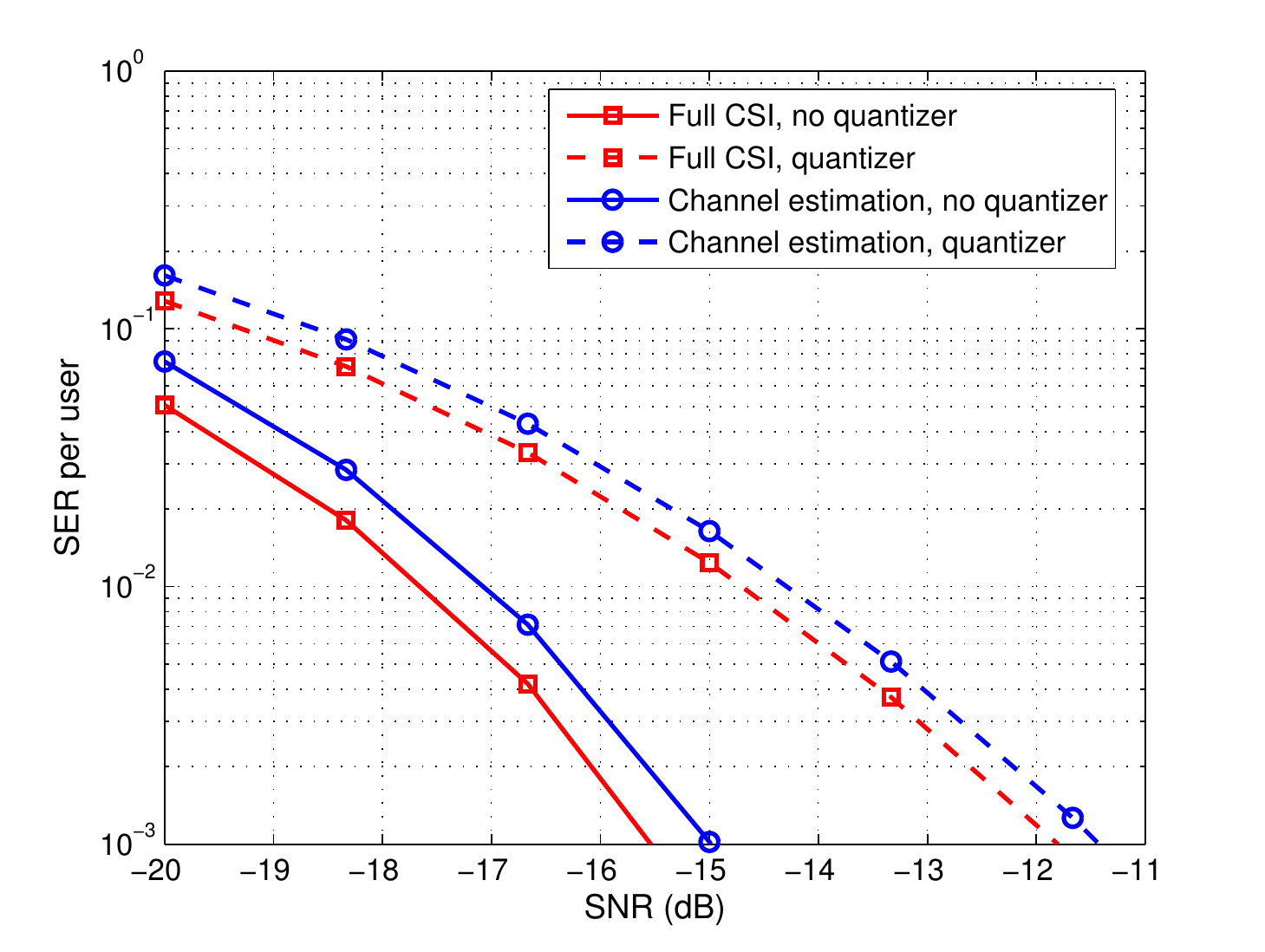} 
\caption{SER  per user versus SNR considering $M=400$ and $K=20$, for the case in which the quantizer is taken into account and the case in which it is not. The ZF is considered for both the cases of perfect and imperfect CSI. } \label{fig:SERvsSNR_gap}
\end{figure}

In Figure \ref{fig:MIvsSNR_gap} and Figure \ref{fig:SERvsSNR_gap}, we show the performance gap between the case in which the quantizer is considered 
in the system model and the case in which it is not considered.
We show the mutual information per user versus the SNR and the  SER per user 
versus the SNR, respectively, for the ZF receiver. Both the cases of full CSI and imperfect CSI are depicted. For the channel estimation phase, a
training sequence of length $N=50K$ has been used. From the figures, we observe that the scenario in which the channel is assumed to be perfectly
known at the receiver and the quantizer effects are neglected gives an upper bound on the performance. We also note that, for values of SNR larger 
than $-10 $ dB, the mutual information achieves the maximum possible QPSK mutual information of 2 bits in all the four considered cases.

\begin{figure}
\centering
\includegraphics[width=0.5\textwidth]{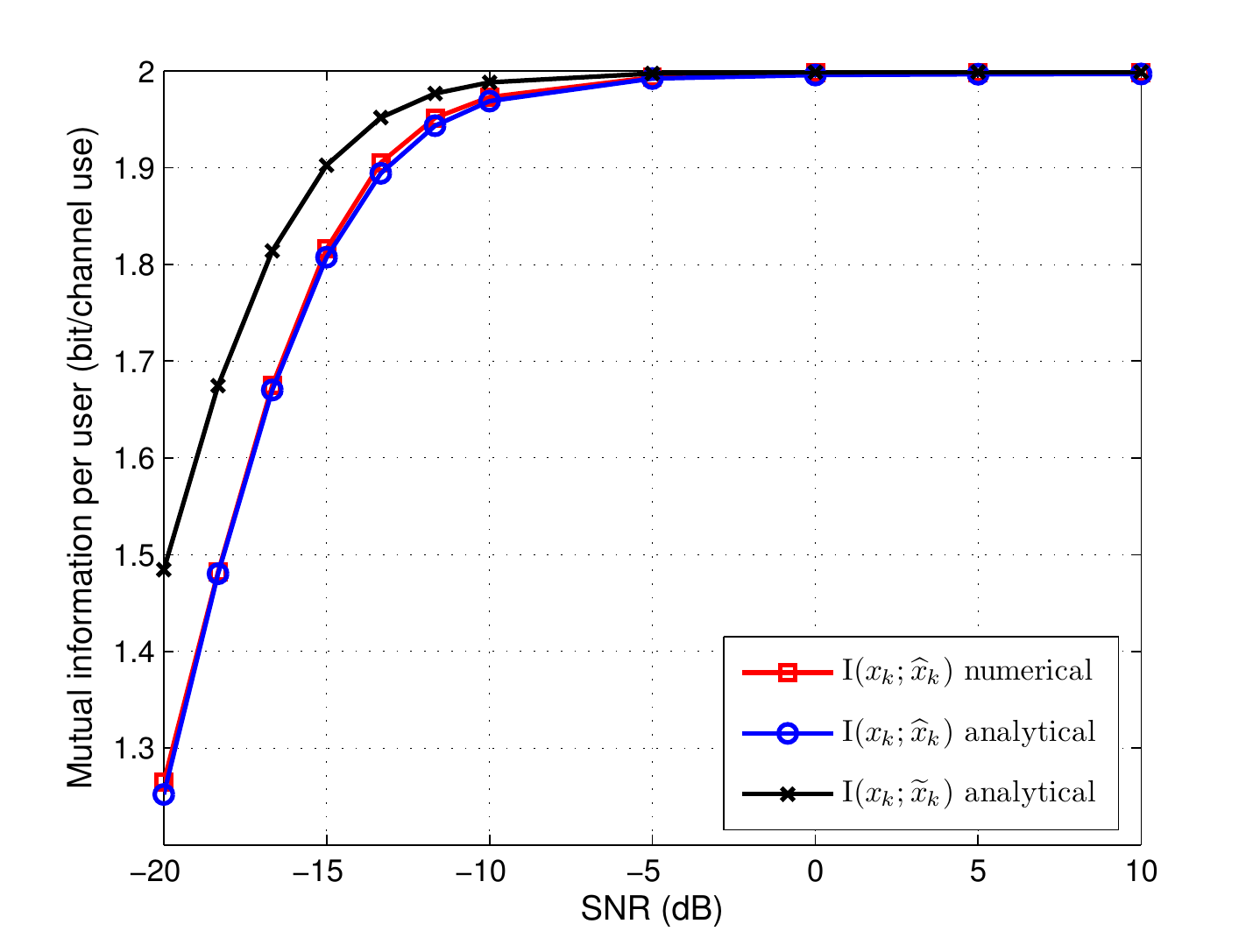} 
\caption{Mutual information per user versus SNR considering $M=400$ and $K=20$, and the MRC filter with full CSI. 
Monte Carlo symbol and noise vectors simulation results are compared with the analytical results
 based on (\ref{pdf_x_soft}).
The mutual information based on (\ref{MI_numerical}) associated with the 
discrete channel between the transmitted and received QPSK symbols $x_k$ and $\widehat{x}_k$,
as well as the mutual information per user between the transmitted QPSK symbol $x_k$ and the soft symbol estimate
$\widetilde{x}_k$ defined as in (\ref{MI_soft}), are 
 depicted.
} \label{fig:MIvsSNR_analytic}
\end{figure}

\begin{figure}
\centering
\includegraphics[width=0.5\textwidth]{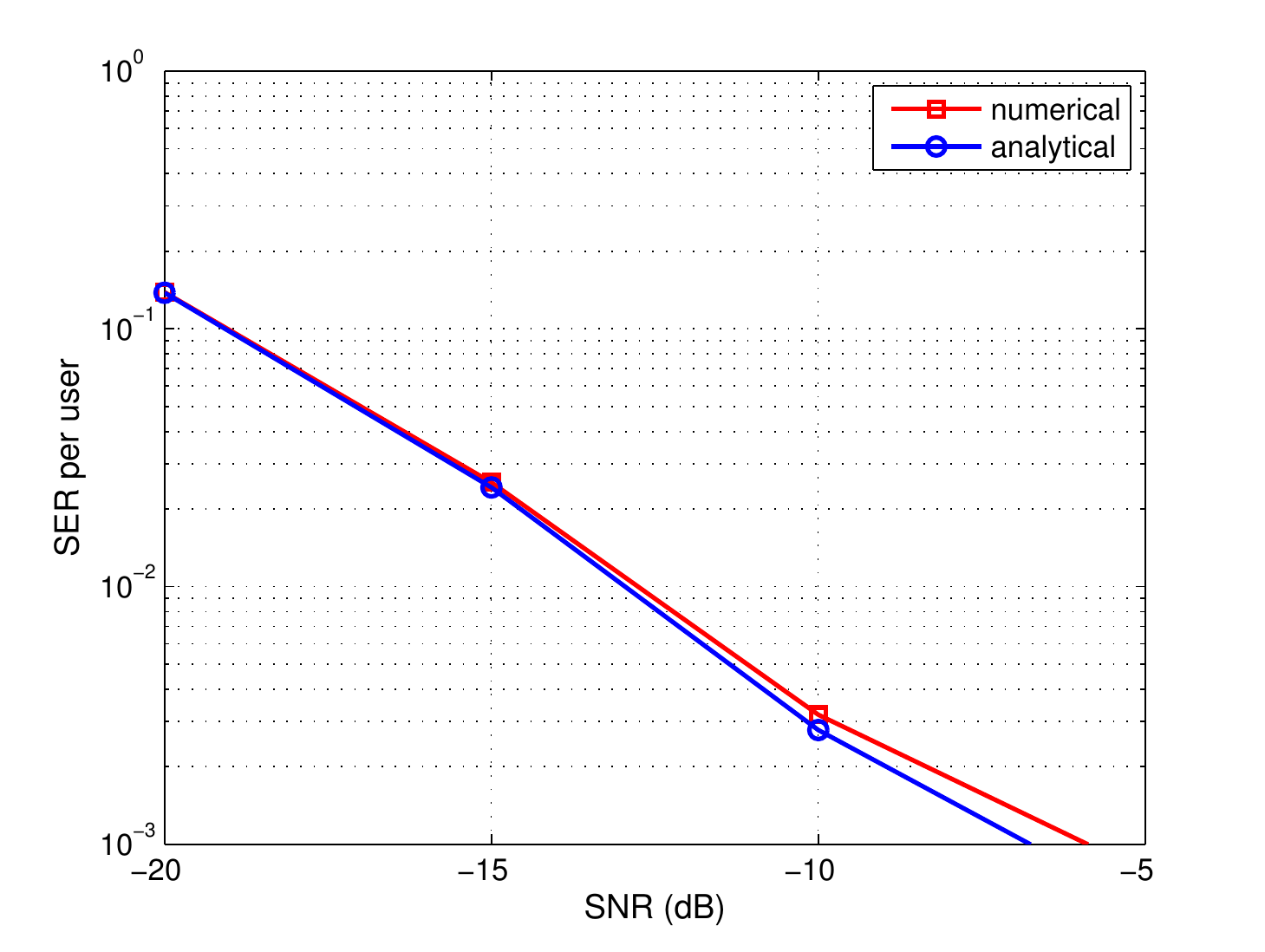} 
\caption{SER  per user versus SNR considering $M=400$ and $K=20$, and the MRC filter with full CSI. Monte Carlo symbol and noise vectors simulation results are compared with the analytical results
 based on (\ref{pdf_x_soft}) and (\ref{SER_analytical}).} \label{fig:SERvsSNR_analytic}
\end{figure}

All the above results have been obtained numerically by symbol and noise vectors Monte Carlo simulations.
In order to verify the analytical analysis in Section \ref{sec:1}, Figure \ref{fig:MIvsSNR_analytic}
and Figure \ref{fig:SERvsSNR_analytic} compare the performance obtained using (\ref{pdf_x_soft}) with those obtained by Monte Carlo
symbol and noise vectors simulations.
In particular, Figure \ref{fig:MIvsSNR_analytic} depicts the mutual information per user versus the SNR, while Figure \ref{fig:SERvsSNR_analytic} 
shows the SER per user versus the SNR. 
For the analytical curves, the transition probabilities ${p(\widehat{x}_k\rvert x_k)}$ are calculated using complementary error functions and (\ref{pdf_x_soft}).
The mutual information based on (\ref{MI_numerical}) associated with the 
discrete channel between the transmitted and received QPSK symbols $x_k$ and $\widehat{x}_k$,
as well as the mutual information per user between the transmitted QPSK symbol $x_k$ and the soft symbol estimate
$\widetilde{x}_k$ defined as in (\ref{MI_soft}), are 
 depicted in Figure \ref{fig:MIvsSNR_analytic}. 
In Figure \ref{fig:SERvsSNR_analytic}, the analytical results are based on the SER expression defined in (\ref{SER_analytical}).
 
The graphs show that the results based on the analytical treatment closely match the symbol and noise vectors Monte Carlo simulation results. 
 As expected in Figure \ref{fig:MIvsSNR_analytic}, the mutual information soft symbol estimate (\ref{MI_soft}) is larger than the 
mutual information in (\ref{MI_numerical}) associated with the discrete channel.


\section{Conclusions}\label{sec:conclusions}
This paper has examined the performance of a massive MIMO uplink system that employs 1-bit ADCs. Numerical evaluation of the mutual information 
and the symbol error rate have been provided for MRC, ZF, and LS receive filters. While the LS filter has been directly calculated depending on the uplink training sequences, the MRC and ZF filters have been derived based on the CSI estimate. We provided a discussion of MAP channel estimation but, due to the the high computational complexity in a setting with many users, we suggested a sub-optimal LS-channel estimation approach.\\
We have also shown how the training sequence length affects 
the performance, and the performance gap between the scenario with  a quantized receive vector and the scenario with an unquantized receive vector.
In general, the ZF filter shows better performance compared to the MRC and the LS filters. However, when the SNR is larger than a certain value,
all the filters achieve the maximal possible QPSK capacity, whatever the training sequence length is, and for both the quantized and the unquantized cases.\\
Further, an analytical analysis of the performance is provided for the MRC filters. We showed that the results based on the analytical 
analysis fit well to the Monte Carlo results. Thanks to the closed-form derivation of the symbol estimate pmf, the computational
complexity is reduced in the sense that a symbol and channel noise vectors Monte Carlo simulation is avoided.
Concluding, we have showed that massive MIMO systems exhibit good performance even when employing 1-bit receive signal quantization. Thus, the ADC
implementation complexity and power consumption can be eliminated.


\vspace*{-2mm}
\bibliographystyle{IEEEtran}
\bibliography{references}


%
%
%

\end{document}